\documentclass{article}
\usepackage{spconf}
\usepackage{amssymb}
\usepackage{graphicx}
\usepackage[cmex10]{amsmath}
\usepackage{subfigure}
\usepackage{times}
\usepackage{comment}
\usepackage{multirow}
\usepackage{ifpdf}
\usepackage{amsfonts}
\usepackage{txfonts}
\usepackage{cite}
\usepackage{listings}
\usepackage{xcolor}
\usepackage{url}
\makeatletter
\def\url@leostyle{%
  \@ifundefined{selectfont}{\def\UrlFont{\sf}}{\def\UrlFont{\small\ttfamily}}}
\makeatother

\title{PERCEPTUAL QUALITY COMPARISON BETWEEN SINGLE-LAYER AND SCALABLE VIDEOS AT THE SAME SPATIAL, TEMPORAL AND AMPLITUDE RESOLUTIONS}
\name{Yuanyi~Xue, Yao~Wang\thanks{The authors would like to thank Dr. Yen-Fu~Ou for his help on this work.}}
\address{Department of Electrical and Computer Engineering\\
        Polytechnic Institute of NYU, Brooklyn, NY 11201, U.S.A\\
        Email:yxue01@students.poly.edu, yao@poly.edu}
\begin{document}
%
\maketitle
\begin{abstract}
In this paper, the perceptual quality difference between scalable and single-layer videos coded at the same spatial, temporal and amplitude resolution (STAR) is investigated through a subjective test using a mobile platform. Three source videos are considered and for each source video single-layer and scalable video are compared at 9 different STARs. We utilize paired comparison methods with and without tie option. Results collected from 10 subjects in the \textit{without ``tie''} option and 6 subjects in the \textit{with ``tie''} option show that there is no significant quality difference between scalable and single-layer video when coded at the same STAR. An analysis of variance (ANOVA) test is also performed to further confirm the finding.
\end{abstract}
\section{Introduction}
\label{sec:intro}

Scalable video coding with spatial, temporal and amplitude scalability offers video servers and clients the flexibility in choosing appropriate video layers according to the network bandwidth and the user perference. Given a bandwidth constraint, the spatial resolution (controlled by frame size), temporal resolution (controlled by frame rate) and amplitude resolution (controlled by quantization parameter), can be adjusted such that the optimal perceptual quality can be achieved. However, scalable coding has not been widely adopted in commercial applications so far because of the complexity of scalable coding and the reduced coding efficiency compared to single layer coding. Most of the existing video streaming architectures uses multiple copies of single layer coded videos at different STAR's, and the system will send a version coded at a particular STAR based on the network condition. It is interesting and useful to see whether there are any quality differences between single-layer and scalable videos coded at the same STAR. In~\cite{yuan_PV10}\cite{yenfu_ivmsp}, we have investigated the impact of STAR on the perceptual quality, and derived a model relating the perceptual quality with the STAR. It will be interesting to see whether the same model is also applicable to non-scalable video.


In this work, we report results from subjective tests that compare the perceived quality between single-layer and scalable video, when coded at the same STAR combination. We design our subjective tests based on the paired comparison methods~\cite{paried_comp}. We conduct the test on a mobile platform with a 4.1-inch WVGA (854$\times$480) touch screen running the Android OS. The remainder of this paper is organized as follows: Section \ref{sec:interface} introduces the test interface, the test video pool and test methodology. Section \ref{sec:result} shows and analyzes the subjective test result. We conclude this work in Section \ref{sec:conclusion}.

\section{Testing interface and methodology}
\label{sec:interface}

\subsection{Testing interface}
The subjective tests are conducted on the TI's Zoom2 mobile development platform equipped with a 4.1-inch WVGA multi-touch screen. Our approach for designing the interface is using the Android's own video playback library (Android SDK), while using Java and XML to control the high-level program flow. For details on the user interface design, please see~\cite{yuan_thesis,yuan_PV10,yenfu_ivmsp}.

\subsection{Test video pool}


\begin{figure*}[!htp]
\centering
  \subfigure[City@4CIF/QP36/30Hz/53rd Frame]{\includegraphics[scale=0.115]{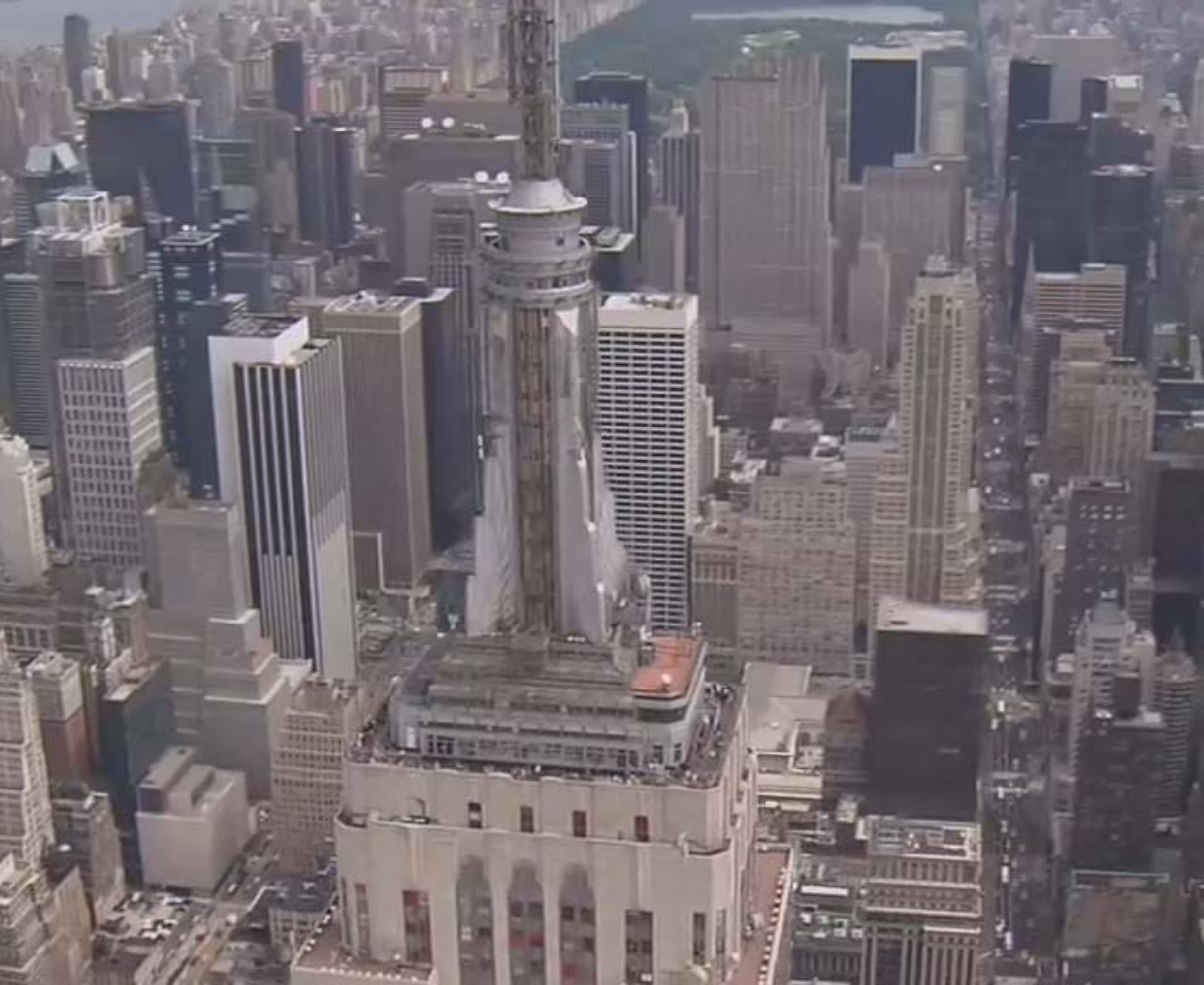}
  \includegraphics[scale=0.115]{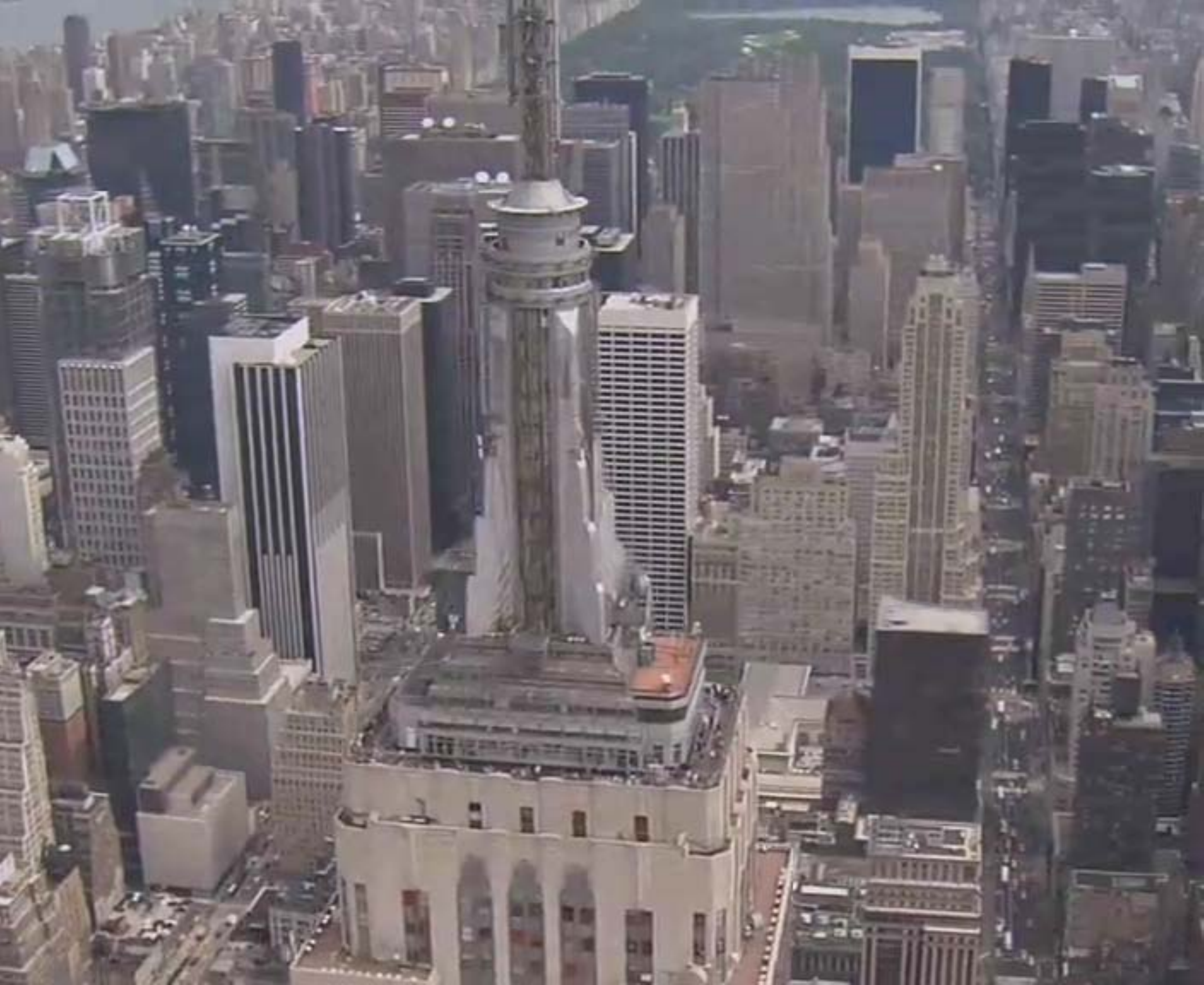}}
  \subfigure[Foreman@CIF/QP28/30Hz/164th Frame]{\includegraphics[scale=0.115]{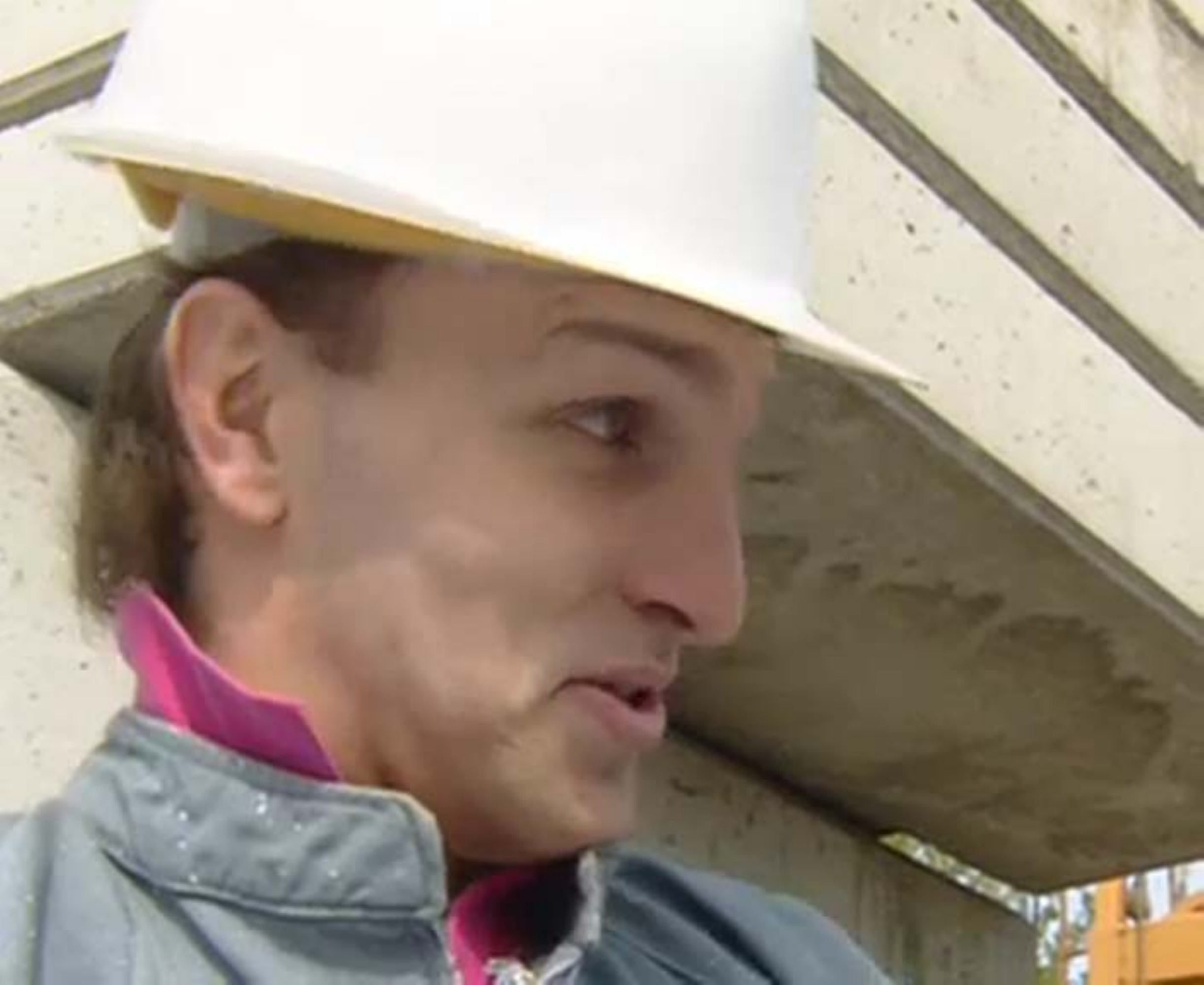}
  \includegraphics[scale=0.115]{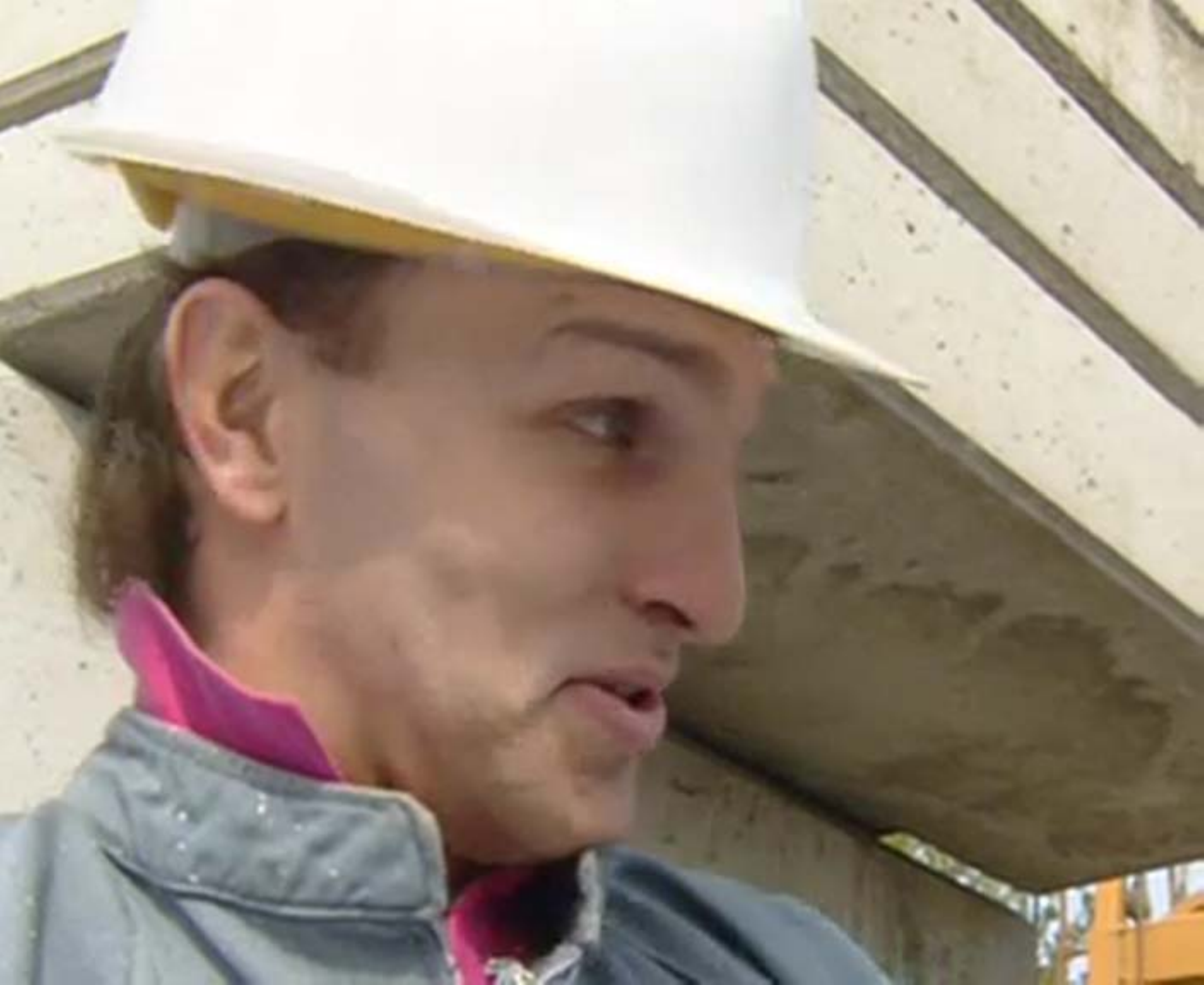}}
  \subfigure[Soccer@4CIF/QP28/15Hz/75th Frame]{\includegraphics[scale=0.115]{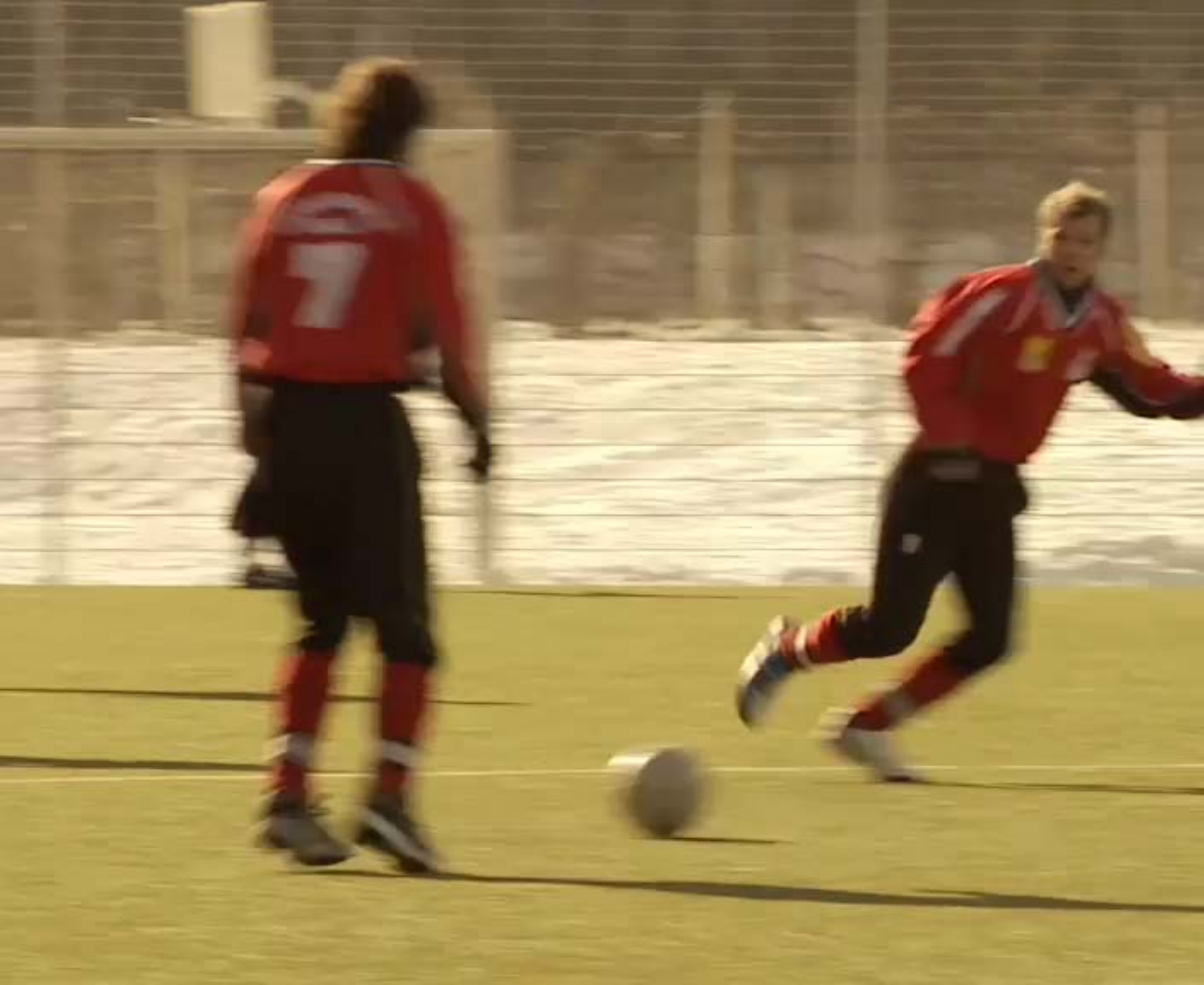}
  \includegraphics[scale=0.115]{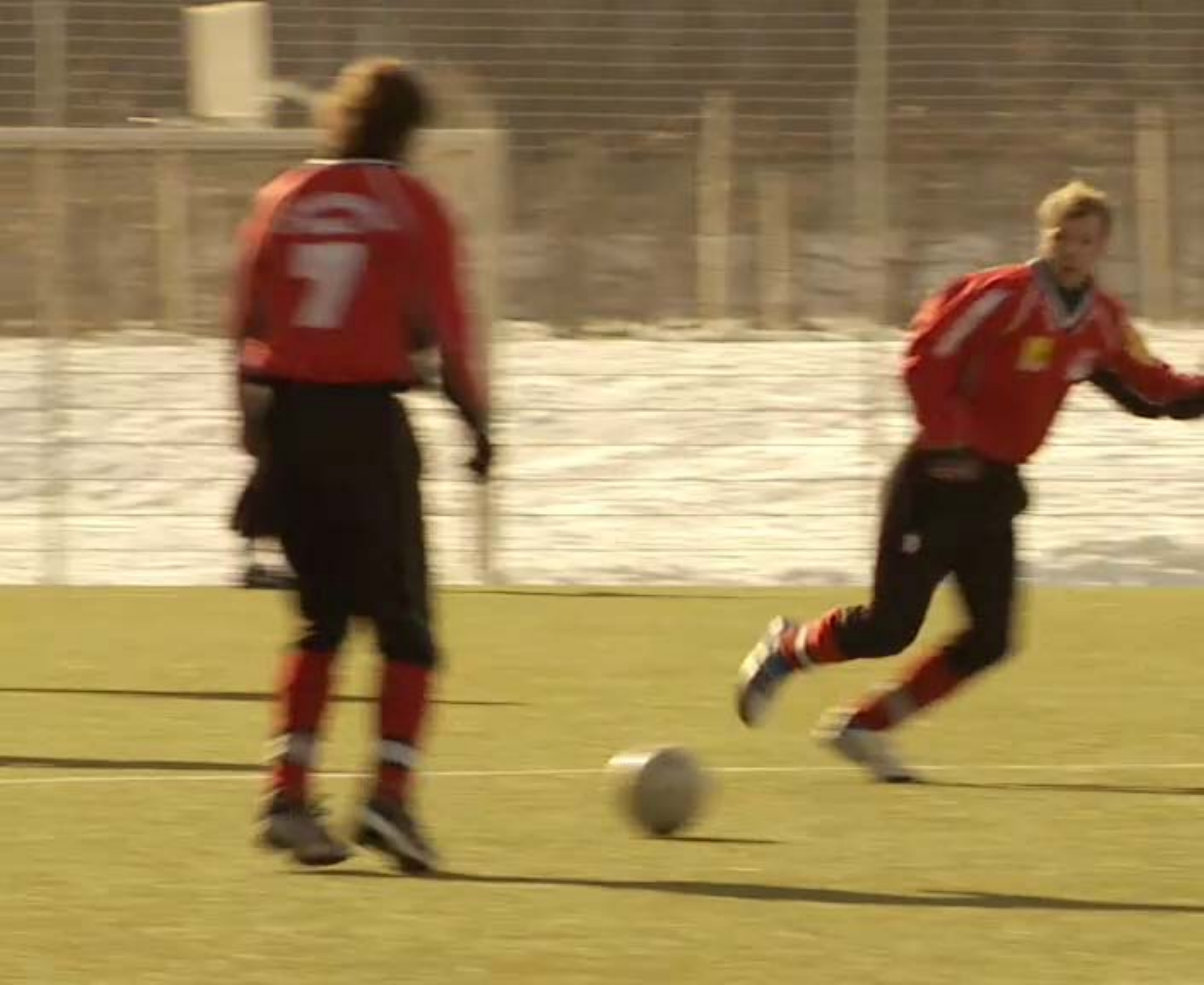}}
  \caption{Comparison snapshots for single-layer and scalable videos at same STAR.}
\label{fig:pvs}
\end{figure*}

Three videos, \emph{city}, \emph{soccer}, and \emph{foreman} from the standard test video database\footnote{Available ftp://ftp.tnt.uni-hannover.de/pub/svc/testsequences/} are used in the test. All videos are originally at 4CIF (704$\times$576) spatial resolution with a frame rate of 30Hz, and each sequence is 8-second long (240 frames). A sine-windowed sinc function, which is the recommended downsampling filter in H.264/SVC standard~ \cite{downsampling}, is used for generating videos at spatial resolutions of CIF and QCIF. The JSVM 9.18~\cite{JSVM} encoder is used to generate both single-layer and layered video. The GOP size is 16 frames in all cases.

We investigate the effect of scalable coding in each dimension (i.e. spatial, temporal, or amplitude scalability) separately, while fixing the resolutions of the other two dimensions at the highest. Specifically, to examine the effect of spatial scalability, we code all videos at highest temporal and amplitude resolution (FR=30Hz, QP=28). To create non-scalable video at different SR's, we code pre-downsampled input videos at QCIF, CIF, and 4CIF resolutions separately, using the JSVM encoder at the single layer mode. To create scalable videos, we create a three layer bitsteam using the JSVM encoder invoking only the spatial scalability, using QCIF as the base layer. For temporal scalability, we fix SR at 4CIF and QP at 28, and produce temporally scalable videos by using the JSVM encoder with the hierarchical B temporal prediction structure, with the base layer corresponding to 3.75 Hz, and additional enhancement layers leading to 7.5, 15, and 30 Hz, respectively. The non-scalable versions are created by coding the pre-downsampled input video at 7.5, 15, and 30 Hz video at the non-scalable mode using the I15BP structure, with only the first frame coded at I mode. Thus, the temporal scalable videos at lower frame rate have higher I/P-frame ratio then the corresponding single-layer videos. No QP cascading is used when temporal and spatial scalability is invoked. Finally, to test amplitude scalability (commonly known as quality or SNR scalability), Coarse Gratitude Scalability (CGS) is used with base layer QP at 44, additional layers using QP at 36 and 28, respectively. For single-layer counterpart, we directly code the video at each QP (to be specific, QP at 28, 36 and 44 individually). Table~\ref{t:TestPts} summarizes the test points examined in different cases.
\begin{table}[htp]
\centering
\caption{Test points}
\label{t:TestPts}
\begin{tabular}{|c|c|}\hline
Common parameters & Test parameters \\ \hline
QP28, 30Hz & 4CIF, CIF and QCIF \\ \hline
4CIF, 30Hz & QP28, QP36 and QP44 \\ \hline
QP28, 4CIF & 30Hz, 15Hz and 7.5Hz \\
\hline
\end{tabular}
\end{table}

The coded bitstream are then extracted and decoded into YUV format, and for CIF and QCIF streams, a 6-tap half-pel with bilinear quarter-pel interpolation filter~\cite{interpolation} is used to upsample it to 4CIF for display in the Zoom2 screen. Finally, single layer and scalable layer videos coded at same STAR are catenated in both ways (single-layer first shown, and scalable-layer first shown) with a 3-second grey ($R=G=B=192$) out interval in between.

\subsection{Methodology}
\label{sec:meth}
To exam whether there is perceptual difference between single-layer and scalable video coded at the same STAR, the paired comparison method~\cite{paried_comp} is used. In paired comparison, a subject views two consecutive videos with a grey-out interval, and then is asked to rate which video is better in terms of perceived quality. There are two approaches in designing subjective tests using paired comparison: 2-forced-choice \textit{without ``tie''} option and 3-forced-choice \textit{with ``tie''} option. In this work, we conduct our subjective tests using both methods. Please remind that for the 2-forced-choice \textit{without ``tie''} test, it is similar to the methodology used in the \textit{just noticeable difference} or jnd test. Here when we count the votes, we are using the $75\%$ jnd criteria.

The subject will view a catenated video from a randomly generated ordering (either single-layer first or scalable first), and after that on the \textit{without ``tie''} test, he/she will choose which one (the first one or the second one) has a better quality; on the \textit{with ``tie''} test, he/she will have the possibility to choose the ``tie'' option if he/she feels the perceived quality is the same for both. The subject can replay the current pair as many times as he/she wishes before rating. For each pair of videos in a particular STAR combination, two occurrences are shown, and the order of which one (single-layer or scalable) shown first is random and determined by the interface. The subject will have to give the opinions (forced choice) on all test points for the session, the total number of test points is 27($3\times3\times3$). Note that with double rating, each subject is viewing and rating 54 nineteen-second sequences.

\section{Result and analysis}
\label{sec:result}

Ten subjects with normal vision participated the 2-forced-choice test, 6 subjects with normal vision participated the 3-forced-choice test. The votes are counted for single-layer and for scalable videos for each test point, respectively.

To provide a intuitive feeling of the PVS, in Fig.~\ref{fig:pvs} we show a set of snapshots of encoded scalable and single-layer videos at the same STAR, and there corresponding scaled absolute difference images. We can see each pair of videos perceptually look very similar, although there are non-zero pixel differences.

\begin{table*}[!htpb]
\centering
{\small
\caption{Votes for 2-forced choice \textit{without ``tie''} option tests}
\label{t:res1}
\scalebox{0.7}{
\begin{tabular}{|c|c|c|c|c|c|c|c|c|} \hline
& \multicolumn{2}{|c|}{\emph{city}} & \multicolumn{2}{|c|}{\emph{soccer}} & \multicolumn{2}{|c|}{\emph{foreman}} & \multicolumn{2}{|c|}{All videos} \\ \hline
& Single & Scalable & Single & Scalable & Single & Scalable & Single & Scalable \\ \hline
4CIF & 13 & 7 & 8 & 12 & 8 & 12 & 29 & 31 \\ \hline
CIF & 8 & 12 & 14 & 6 & 12 & 8 & 34 & 26 \\ \hline
QCIF & 9 & 11 & 10 & 10 & 11 & 9 & 30 & 30 \\ \hline
\textbf{All S} & \textbf{30} & \textbf{30} &
\textbf{32} & \textbf{28} & \textbf{31} & \textbf{29} & \textbf{93} & \textbf{87} \\ \hline
30Hz & 9 & 11 & 12 & 8 & 10 & 10 & 31 & 29 \\ \hline
15Hz & 13 & 7 & 9 & 11 & 11 & 9 & 33 & 27 \\ \hline
7.5Hz & 6 & 14 & 7 & 13 & 12 & 8 & 25 & 35 \\ \hline
\textbf{All T} & \textbf{28} & \textbf{32} & \textbf{28} & \textbf{32} & \textbf{33} & \textbf{27} & \textbf{89} & \textbf{91} \\ \hline
QP28 & 10 & 10 & 11 & 9 & 10 & 10 & 31 & 29\\ \hline
QP36 & 8 & 12 & 13 & 7 & 12 & 8 & 33 & 27 \\ \hline
QP44 & 5 & 15 & 11 & 9 & 14 & 6 & 30 & 30 \\ \hline
\textbf{All Q} & \textbf{23} & \textbf{37} & \textbf{35} & \textbf{25} & \textbf{36} & \textbf{24} & \textbf{94} & \textbf{86} \\ \hline
\end{tabular}}}
\end{table*}

\begin{figure}
\label{f:anova}
\centering
\includegraphics[width=5cm]{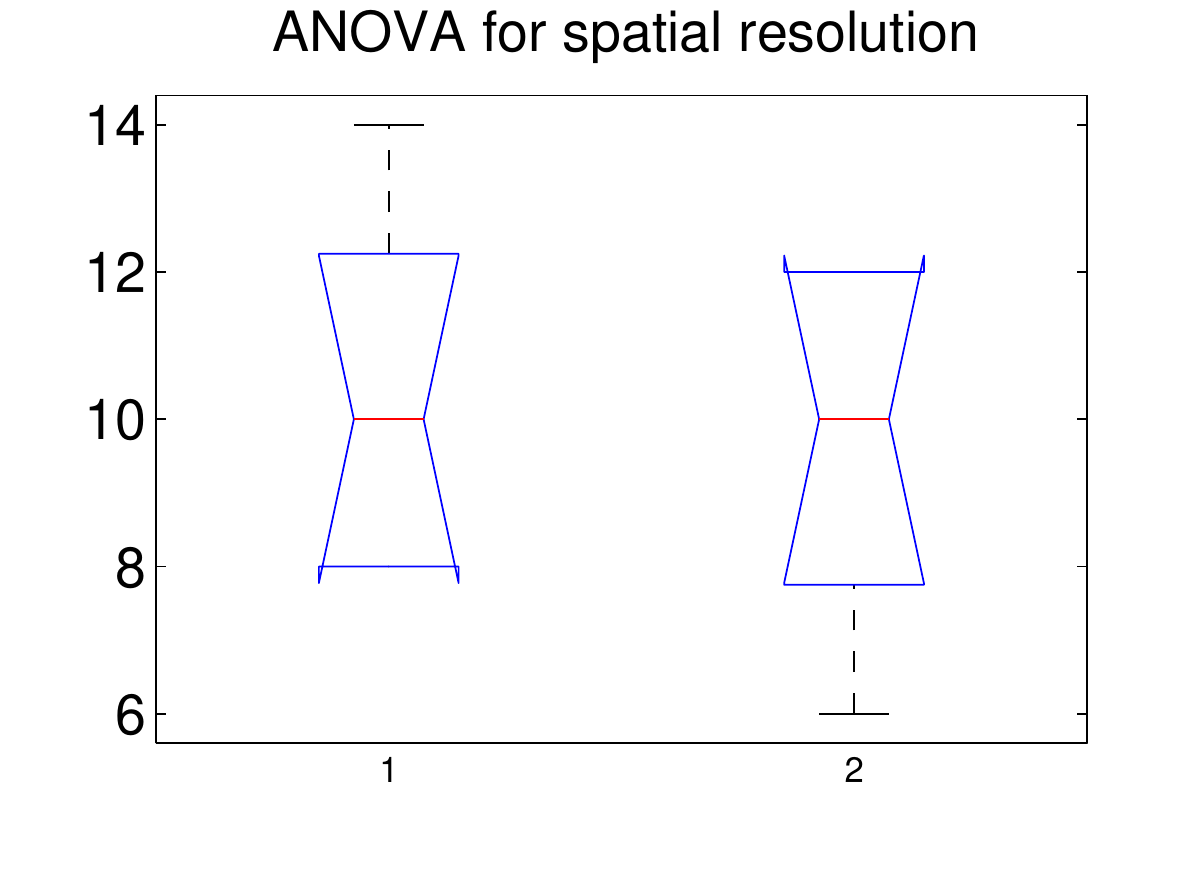}
\includegraphics[width=5cm]{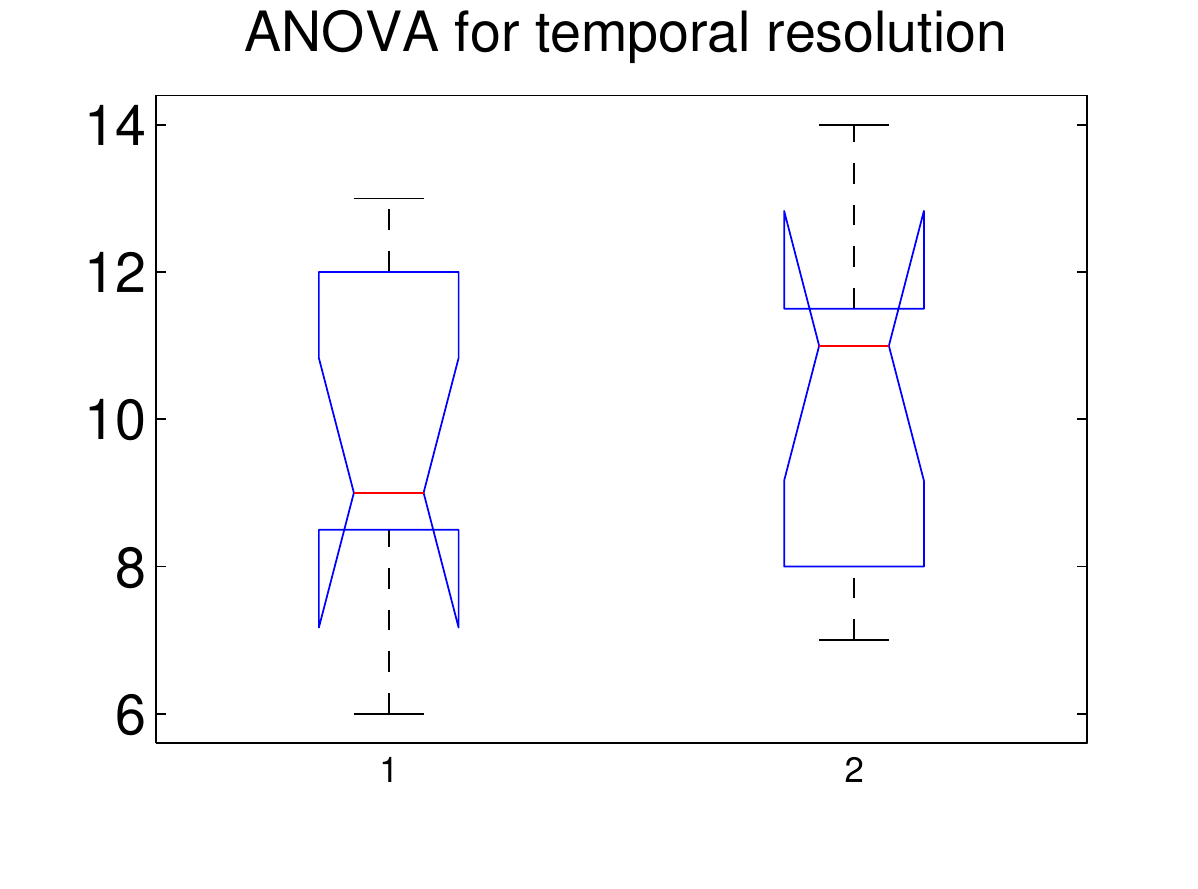}
\includegraphics[width=5cm]{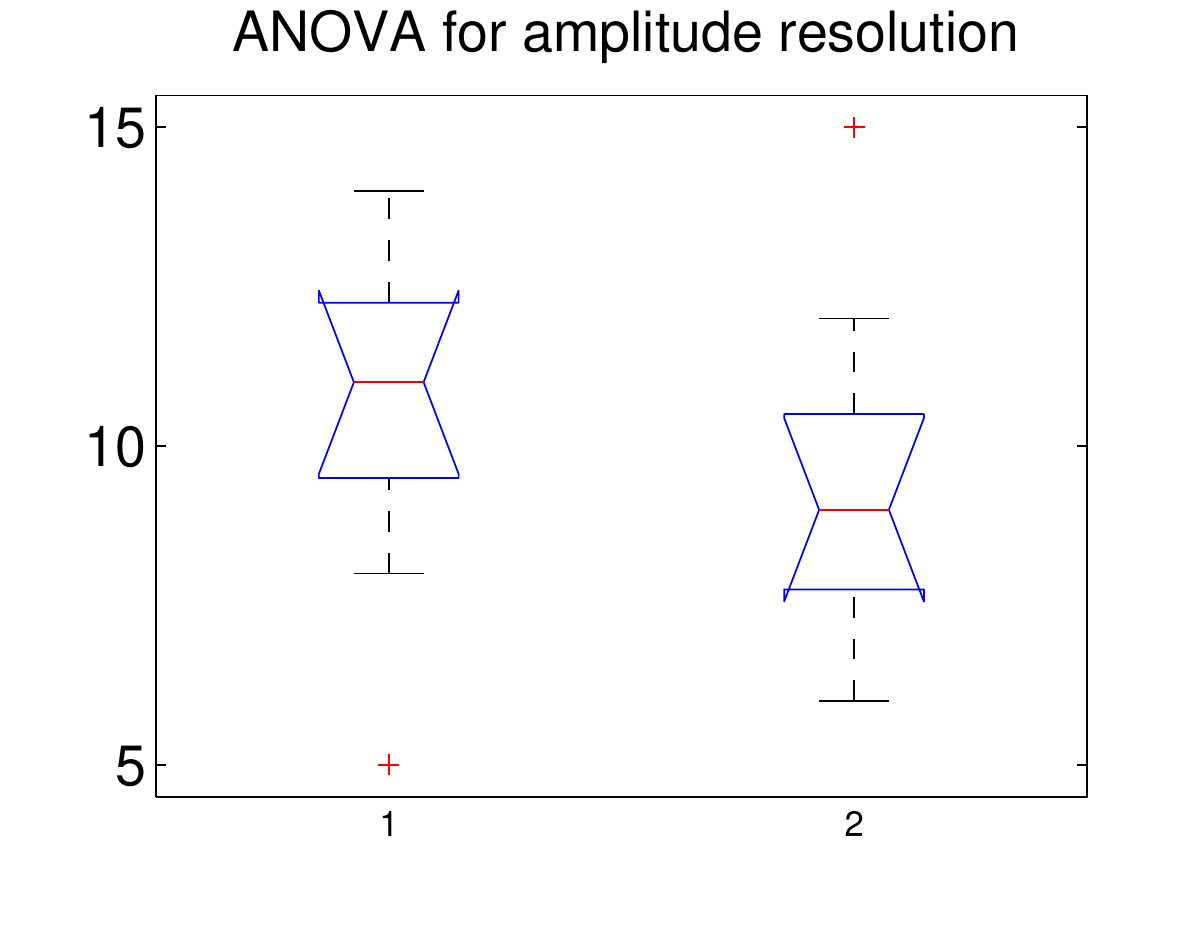}
\caption{Box plots for the ANOVA tests, in x-axis, 1 indicates single layer video and 2 indicates scalable video.}
\end{figure}
\begin{table}[htp]
\centering
\caption{$p$-value and $f$-value of ANOVA test for \textit{without ``tie''} test}
\label{t:anova}
\begin{tabular}{|c|c|c|} \hline
 & $p$-value & $f$-value \\ \hline
Spatial & $0.5458$ & $0.38$ \\ \hline
Temporal & $0.5549$ & $0.36$ \\ \hline
Amplitude & $0.4946$ & $0.49$ \\ \hline
\end{tabular}
\end{table}

\begin{figure}[!htp]
\centering
  {\subfigure[MAD=1.9299]{\includegraphics[width=4.1cm]{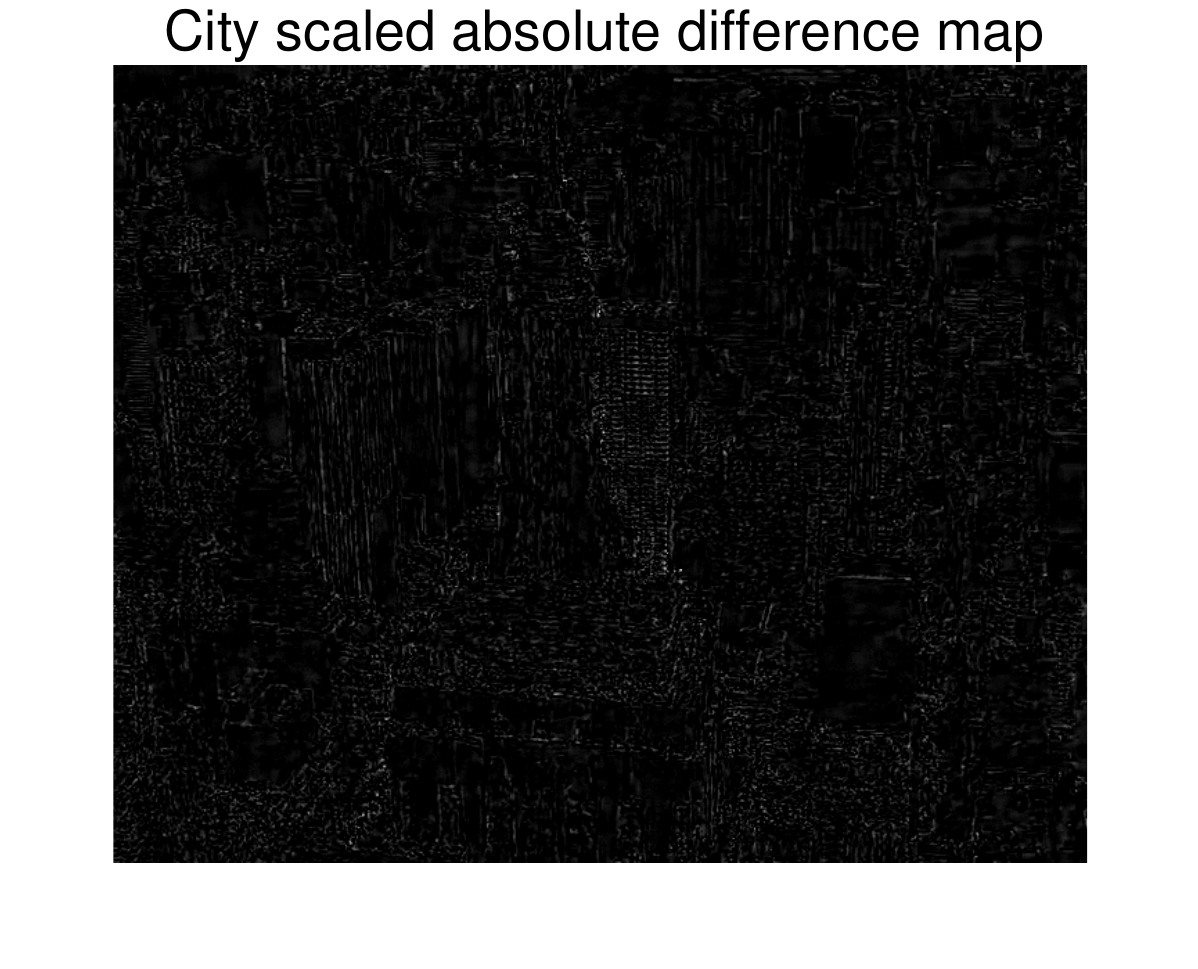}}
  \subfigure[MAD=0.9875]{\includegraphics[width=4.1cm]{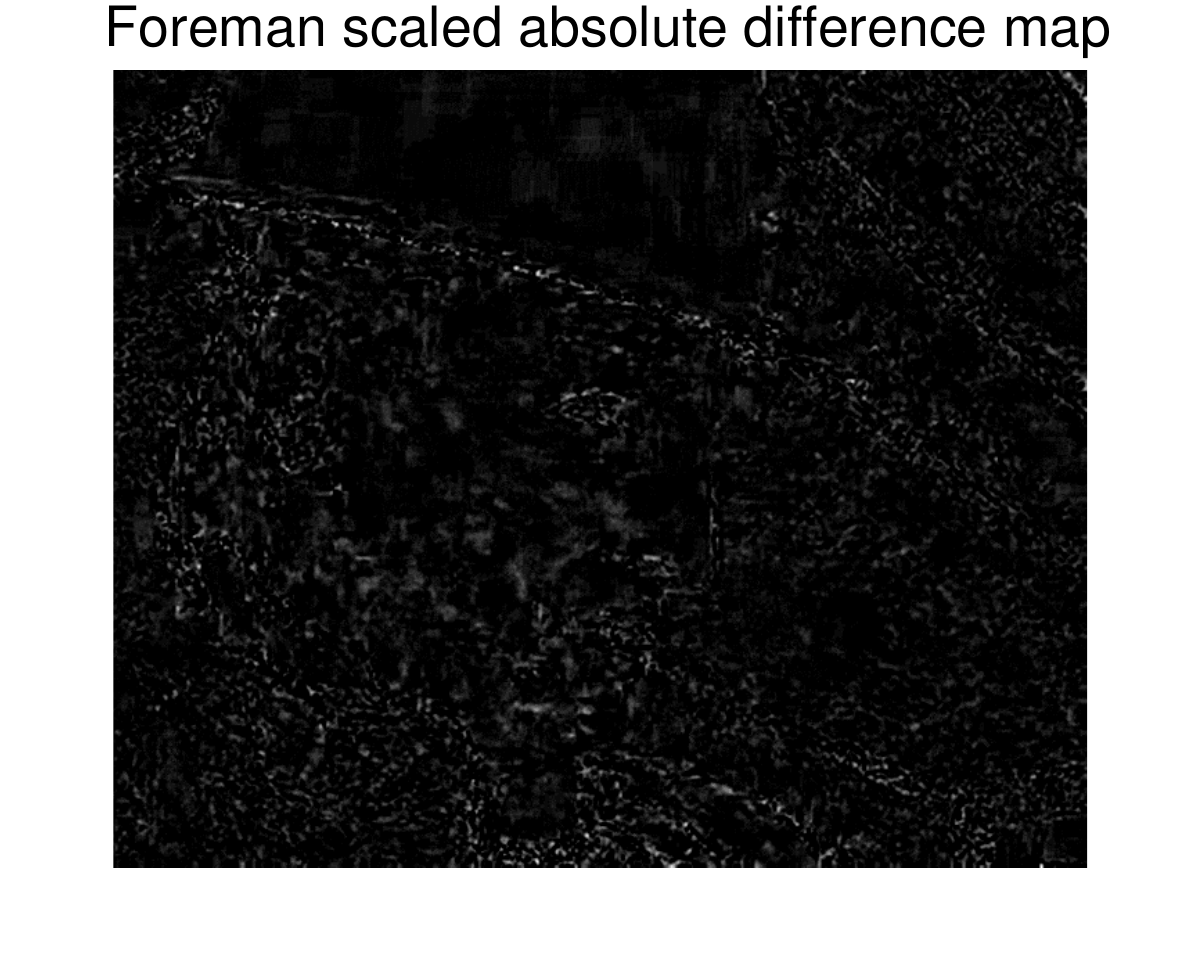}}
  \subfigure[MAD=1.2610]{\includegraphics[width=4.1cm]{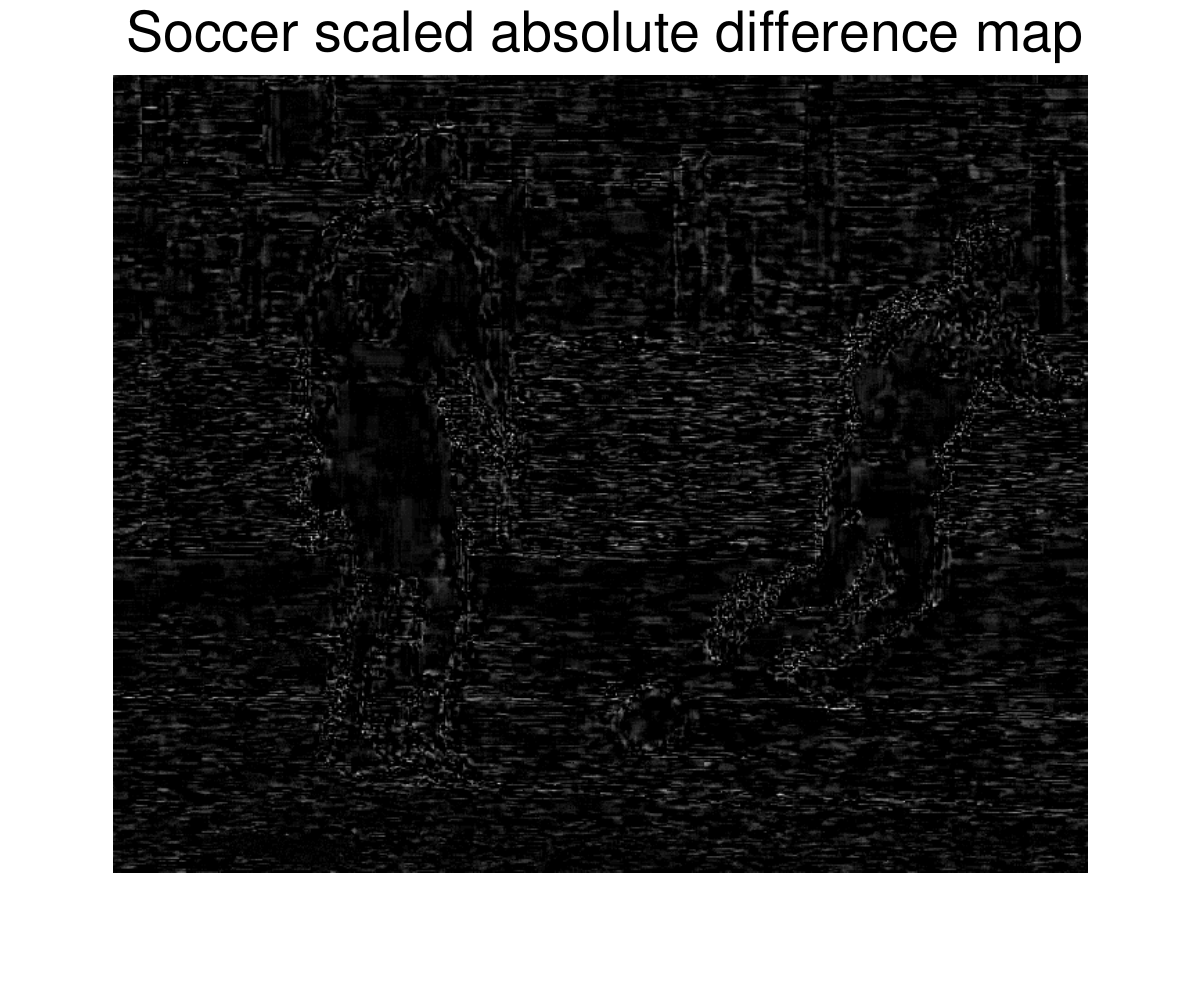}}}
  \label{fig:pvs2}
  \caption{Scaled absolute difference maps for three videos}
\end{figure}

\begin{table*}[!htpb]
\centering
{\small
\caption{Votes for 3-forced choice \textit{with ``tie''} option tests}
\label{t:res2}
\scalebox{0.7}{
\begin{tabular}{|c|c|c|c|c|c|c|c|c|c|c|c|c|} \hline
& \multicolumn{3}{|c|}{\emph{city}} & \multicolumn{3}{|c|}{\emph{soccer}} & \multicolumn{3}{|c|}{\emph{foreman}} & \multicolumn{3}{|c|}{All videos} \\ \hline
& Single & Scalable & Tie & Single & Scalable & Tie & Single & Scalable & Tie & Single & Scalable & Tie \\ \hline
4CIF & 1 & 1 & 10 & 0 & 2 & 10 & 1 & 3 & 8 & 2 & 6 & 28\\ \hline
CIF & 1 & 2 & 9 & 2 & 1 & 9 & 0 & 2 & 10 & 3 & 5 & 28\\ \hline
QCIF & 2 & 2 & 8 & 1 & 0 & 11 & 0 & 0 & 12 & 3 & 2 & 31\\ \hline
\textbf{All S} & \textbf{4} & \textbf{5} &
\textbf{27} & \textbf{3} & \textbf{3} & \textbf{30} & \textbf{1} & \textbf{5} & \textbf{30} & \textbf{8} &\textbf{13} & \textbf{87} \\ \hline
30Hz & 2 & 2 & 8 & 1 & 3 & 8 & 2 & 1 & 9 & 5 & 6 & 25 \\ \hline
15Hz & 3 & 1 & 8 & 2 & 3 & 7 & 3 & 2 & 7 & 8 & 6 & 22 \\ \hline
7.5Hz & 2 & 2 & 8 & 1 & 1 & 10 & 2 & 3 & 7 & 5 & 6 & 25 \\ \hline
\textbf{All T} & \textbf{7} & \textbf{5} & \textbf{24} & \textbf{4} & \textbf{7} & \textbf{25} & \textbf{7} & \textbf{6} & \textbf{23} & \textbf{18} & \textbf{18} & \textbf{72} \\ \hline
QP28 & 1 & 2 & 9 & 2 & 3 & 7 & 2 & 2 & 8 & 5 & 7 & 24 \\ \hline
QP36 & 2 & 2 & 8 & 1 & 3 & 8 & 0 & 1 & 11 & 3 & 6 & 27 \\ \hline
QP44 & 1 & 1 & 10 & 2 & 2 & 8 & 1 & 0 & 11 & 4 & 3 & 29 \\ \hline
\textbf{All Q} & \textbf{4} & \textbf{5} & \textbf{27} & \textbf{5} & \textbf{8} & \textbf{23} & \textbf{3} & \textbf{3} & \textbf{30} & \textbf{12} & \textbf{16} & \textbf{80} \\ \hline
\end{tabular}}}
\end{table*}

Table~\ref{t:res1} provides the counting result for the 2-forced-choice test. As we mentioned in Section \ref{sec:meth}, the 2-forced-choice test can be seen as a special case of JND test. If the hypothesis that there is a ``\textit{just noticeable}'' difference on the perceived quality is accepted, the winning frequency for the better quality one should be at least above $75\%$ under the $75\%$ jnd condition, that is at least 15 votes for a particular video at a particular STAR combination, since each video pair is viewed 20 times. From Table~\ref{t:res1}, except \emph{city} at QP44/30Hz/4CIF, there is no such occurrence. Thus it's safe to say that there is no significant difference in the perceptual quality between the scalable and single-layer video at all STAR's examined.

To further examine the statistical significance of the rating differences, we conducted an ANOVA analysis in the three dimensions separately and the results are shown in Table~\ref{t:anova}. For all the cases, the $p$-values are larger than $0.05$, indicating that there are no significant differences between videos coded in single-layer and scalable modes. We also show the box plots of the ANOVA tests in Fig.~\ref{f:anova}. In the box plots, the central red mark is the median of the data, the notches in the box represent the 95\% confidence interval of the median, the edges of the box are the 25th and 75th percentiles and the whiskers extend to the most extreme data points. We find that the $95\%$ confidence interval of medians are overlapped, indicating there is no perceived quality difference between single-layer and scalable coded videos.

Table~\ref{t:res2} shows the counting result for the 3-forced-choice test. We see that in most cases, the majority of votes are given to the ``tie'' option, indicating the viewers could not tell the difference between the single-layer and scalable coded video at the same STAR.
%
%
\section{Conclusion}
\label{sec:conclusion}
This paper reports results from a perceptual quality assessment comparing single-layer video and scalable video, when coded at the same spatial, temporal and amplitude resolutions (STAR). The subjective test was conducted using paired comparison with and without ``tie'' option and double rating. Ten subjects' data were collected for the \textit{without ``tie''} option, and 6 subjects' ratings for the \textit{with ``tie''} option. The test result shows that under the same STAR there is no significant perceptual quality difference between single layer coded video and scalable one, both by observing the ratings and through using the ANOVA test. Although the single-layer and scalable videos are generated using the H.264/AVC and H.264/SVC compliant codecs, respectively (both implemented via the JSVM encoder under different settings), we believe the conclusion may be generally true for any videos coded at the same STAR, regardless the encoding method. Note here we measure the amplitude resolution by the inverse of quantization stepsize. We consider the two videos as having the same amplitude resolution if they are quantized using the same type of quantizer and at the same quantization stepsize. One important consequence of our finding here is that the Q-STAR model developed in our prior work~\cite{yenfu_ivmsp} modeling the perceptual quality as a function of STAR is applicable to both scalable and non-scalable video.

\bibliographystyle{IEEEbib}
{\small
\bibliography{ICIP2012}}

\end{document}